\RequirePackage{amsmath}
\documentclass{llncs}

\usepackage[utf8]{inputenc}
\usepackage{amsmath,amssymb}
\usepackage{times,helvet,courier}
\usepackage{tikz}
\usetikzlibrary{arrows,automata,shapes.geometric}
\usepackage{array}
\usepackage{listings}
\usepackage{refcount}

\newcommand{\naf}[1]{\ensuremath{{\mathit{not}\;{#1}}}} 

\newcommand{\goLeft}[1]{\ensuremath{\mathit{left}(#1)}}
\newcommand{\goRight}[1]{\ensuremath{\mathit{right}(#1)}}
\newcommand{\Out}[1]{\ensuremath{\mathit{out}(#1)}}
\newcommand{\At}[1]{\ensuremath{\mathit{at}(#1)}}
\newcommand{\Ab}[1]{\ensuremath{\mathit{ab}(#1)}}
\newcommand{\Abd}[1]{\ensuremath{\mathit{ab'}(#1)}}

\newcommand{\Table}[1]{\ensuremath{\mathit{table}(#1)}}
\newcommand{\liftLeft}{\ensuremath{\mathit{lift\_l}}}
\newcommand{\liftRight}{\ensuremath{\mathit{lift\_r}}}
 
\title{Formalizing Multi-Agent Systems Using Action Descriptions in Single Agent Perspective}

\author{Orkunt Sabuncu\inst{1,2} \ Torsten Schaub\inst{1} \ Christian Schulz-Hanke\inst{1}}

\institute{University of Potsdam, Germany \and NOVA LINCS, Universidade Nova de Lisboa, Portugal}

\begin{document}

\maketitle

\begin{abstract}
Logic-based representations of multi-agent systems 
have been extensively studied.
In this work, we focus on the action language $\mathcal{BC}$ to formalize global views of MAS domains.
Methodologically,
we start representing the behaviour of each agent by an action description from a single agent perspective.
Then,
it goes through two stages that guide the modeler in composing the global view by 
first designating multi-agent aspects of the domain via potential conflicts and 
later resolving these conflicts according to the expected behaviour of the overall system.
Considering that representing single agent descriptions is relatively simpler than representing multi-agent description directly,
the formalization developed here is valuable from a knowledge representation perspective.
\end{abstract}


\section{Introduction}\label{sec:introduction}

Logic-based representations of multi-agent systems (MAS)
have been extensively studied.
Well-defined semantics of such representations allow for carrying out various reasoning tasks about MAS.
Action languages \cite{gellif98a}
are among these representations of MAS.
They have been successfully applied to represent and reason not only
on single agent \cite{bargel00a,balgel08a}
but also multi-agent domains \cite{baral97a,basopo09a,gelwat07a}.

In this work, we focus on the action language $\mathcal{BC}$ \cite{leliya13a} to formalize MAS domains.
More specifically, the formalization deals with the global view of MAS domains,
which captures all knowledge about environment and capabilities of agents.
It is basically modelled as a transition diagram whose vertices and edges denote states of the domain 
and sets of agents' actions occurring concurrently, respectively.

Unlike earlier works \cite{basopo09a,gelwat07a,chgewa07a}, 
where the modeler is expected to encode the global view using an action language from scratch,
our formalization takes action descriptions of each agent from a single agent perspective as input 
and goes through two stages focusing on different aspects of the domain.
Note that representing an action description of an agent with a local perspective is relatively easier than 
representing the whole multi-agent domain directly due to interacting concurrent actions.
The first stage in our methodology establishes an intermediate action description
that covers all valid states of the global view and 
identifies cases of the domain where actions of agent interact and have effects that are invisible from a single agent perspective.
To this end,
we use \emph{potential conflict} to specify such cases formally.
The second stage addresses identified potential conflicts by resolving them so that
corresponding concurrent actions have the desired effect in the environment of the domain.
The resolution is achieved via defeating involved dynamic laws of the intermediate description.
This is possible via a syntactic transformation that makes dynamic laws of an action description defeasible.
Formal properties of this transformation are given and
the existence of a resolution at the second stage is proved.
The resulting action description represents the transition diagram corresponding to the global view of the MAS domain.

The action language $\mathcal{BC}$ can express defeasible laws.
This capacity is crucial for our formalization and has been our main motivation in selecting $\mathcal{BC}$ 
as the underlying action language.

The formalization,
which is the novel contribution of this work,
guides the modeler and structures her efforts by making multi-agent aspects of the domain explicit.
To the best of our knowledge, such a methodology has not been explicitly defined and employed before.

While the scope of this work focuses on the global view of a MAS domain,
it is important to emphasize that real agents residing in the environment may have autonomy via their own local view 
exhibiting a behaviour different than the one represented by the agent description from a single agent perspective used in the methodology.
In fact they can utilize their own strategies on how to behave and even try to perform illegal actions.
It is up to the MAS architecture utilizing this global view to react to these illegal actions 
(by simply discarding them or issuing an execution failure).

In what follows, 
we provide background information on the action language $\mathcal{BC}$.
We also illustrate an action description describing a single Sumo agent,
which is the base of the running example used throughout the paper.
In Section~\ref{sec:foundation},
we describe our methodology of formalizing MAS with two stages.
Finally,
we conclude the paper with discussion that also includes future lines of research in Section~\ref{sec:discussion}.


\section{Background}\label{sec:background}

We give below syntax and semantics of the Boolean fragment of slightly modified action language $\mathcal{BC}$, 
as introduced in~\cite{leliya13a}.

We consider a signature of Boolean \emph{action} and \emph{fluent symbols},
denoted by $\mathcal{A}$ and $\mathcal{F}$, respectively.
In what follows, we simply refer to them as actions and fluents. 
Fluents are further divided into \emph{regular} and \emph{defined} fluents.
A defined fluent is useful for representing a property that is statically determined in terms of other fluents.
A \emph{fluent literal} is a fluent $a$ or its negation $\neg a$.
Similarly, an \emph{action literal} is an action $c$ or its negation $\neg c$.
A \emph{static law} is an expression of the form
\begin{equation}\label{eq:static:law}
a_0\textbf{ if }a_1,\dots,a_m\textbf{ ifcons } a_{m+1},\dots,a_n
\end{equation}
where $a_i$ is a fluent literal for $0\leq m\leq n$;
a \emph{dynamic law} is an expression of the form
\begin{equation}\label{eq:dynamic:law}
a_0\textbf{ after }a_1,\dots,a_m\textbf{ ifcons } a_{m+1},\dots,a_n
\end{equation}
where
$a_0$ is a regular fluent literal,
each of $a_1,\dots,a_m$ is a fluent or action literal,
and
$a_{m+1},\dots,a_n$ are fluent literals.
An \emph{action description} is a finite set of of static and dynamic laws.
The ifcons part of a law is important for representing defaults.
Various abbreviations, 
such as $\textbf{impossible}, \textbf{nonexecutable}, \textbf{inertial},$ and $\textbf{default}$ laws,
are useful for developing succinct action descriptions.
They can be rewritten in terms of static and dynamic laws (refer to~\cite{leliya13a} for their definitions).

The semantics of action descriptions is given in terms of transition systems induced by 
a translation into logic programs under stable models semantics~\cite{gellif88b}.
To be more precise,
an action description $\mathcal{D}$ and a horizon $l$
yield a program $P_l(\mathcal{D})$ whose 
stable models represent all paths of length $l$ in the transition system corresponding to $\mathcal{D}$.
The signature of each program $P_l(\mathcal{D})$ consists of labeled expressions of the form
\(
i:a
\),
where $i\leq l$ and $a$ is a fluent literal, or 
$i< l$ and $a$ is an action literal.
As defined in~\cite{leliya13a}, each program $P_l(\mathcal{D})$ consists of the following rules
\begin{enumerate}
\item for each static law of form \eqref{eq:static:law} in $\mathcal{D}$ and $i\leq l$ a rule of form
\begin{equation}\label{eq:static:rule}
i:a_0\leftarrow i:a_1,\dots,i:a_m,\naf{\naf{i:a_{m+1}}},\dots,\naf{\naf{i:a_n}}
\end{equation}
\item for each dynamic law of form \eqref{eq:dynamic:law} in $\mathcal{D}$ and $i< l$ a rule of form
\begin{equation}\label{eq:dynamic:rule}
(i+1):a_0\leftarrow i:a_1,\dots,i:a_m,\naf{\naf{(i+1):a_{m+1}}},\dots,\naf{\naf{(i+1):a_n}}
\end{equation}
\item for each regular fluent $f$ in $\mathcal{D}$ a choice rule of form
\begin{equation}\label{eq:regular:choice}
\{0:f,0:\neg f\}
\end{equation}
\item for each action $a$ in $\mathcal{D}$ and $i< l$ a choice rule of form
\begin{equation}\label{eq:action:choice}
\{i:a\}
\end{equation}
\item for each fluent $f$ in $\mathcal{D}$ and $i\leq l$ an integrity constraint of form
\begin{equation}\label{eq:fluent:value:uniqueness}
\leftarrow\{i:f, i:{\neg f}\}\neq 1
\end{equation}
\item for each action $a$ in $\mathcal{D}$ and $i< l$ a rule of form
\begin{equation}
\label{eq:action:negative}
i:\neg a \leftarrow \naf{i:a}
\end{equation}
\end{enumerate}    
For a set $X$ of labeled expressions and $i\geq 0$, define
\(
X|_i=\{a\mid i:a\in X\}
\)
The transition system $(S(\mathcal{D}),T(\mathcal{D}))$ induced by an action description $\mathcal{D}$
is then defined as follows.
\footnote{Note that $X|_1=X|_1\cap\mathcal{L}$.}
\begin{align}
S(\mathcal{D})
&=
\{X|_0\mid X \textit{ is a stable model of }P_0(\mathcal{D})\}
\\
T(\mathcal{D})
&=
\{\langle X|_0\cap\mathcal{L}, X|_0\cap\mathcal{A}, X|_1\rangle\mid X \textit{ is a stable model of }P_1(\mathcal{D})\}
\end{align}
where $\mathcal{L}=\mathcal{F}\cup\{\neg f\mid f\in\mathcal{F}\}$

Note that unlike the logic program used for semantics of $\mathcal{BC}$ in  \cite{leliya13a} 
we name unperformed actions by explicitly generating negative action literals in \eqref{eq:action:negative}
and consequently allow negative action literals in the after part of a dynamic law.
This is critical in a multi-agent setting to represent situations where an agent does not perform a specific action.
Such a case is illustrated using the running example in Section~\ref{sec:foundation}.

\label{sumo_example}
For illustration, consider a (single) Sumo agent in a ring divided into $l$ horizontal slots.
A Sumo can move left or right to adjacent slots in the ring,
and may drop out at each end.
We capture this through 
fluents $\At{A,L}$, $\Out{A}$ and 
actions \goLeft{A}, \goRight{A}, respectively,
where
the variable $A$ stands for an agent identifier\footnote{Strictly speaking, 
  agent identifiers are obsolete in a single-agent environment 
  but their introduction paves the way for the multi-agent setting in the next section.}
and $L,L' \in \{ 1,\dots,l \}$.

The behavior of our simple Sumo agent in an $l$ slot ring is represented by the following action description,
composed of static laws \eqref{sumo:static:1}--\eqref{sumo:static:4} 
and dynamic laws \eqref{sumo:dynamic:1}--\eqref{sumo:dynamic:7}, respectively.
\begin{align}
\label{sumo:static:1}&\neg \At{A,L} \textbf{ if } \At{A,L'} & (L \neq L')\\ 
\label{sumo:static:2}&\neg \Out{A} \textbf{ if } \At{A,L} & \\
\label{sumo:static:3}&\neg \At{A,L} \textbf{ if } \Out{A} & \\
\label{sumo:static:4}&\textbf{impossible } \neg \At{A,1}, \dots, \neg \At{A,l}, \neg \Out{A} & \\  
\label{sumo:dynamic:1}&\At{A,L} \textbf{ after } \goLeft{A}, \At{A,L'} & (L=L'-1)  \\
\label{sumo:dynamic:2}&\At{A,L} \textbf{ after } \goRight{A}, \At{A,L'} &  (L=L'+1)  \\
\label{sumo:dynamic:3}&\Out{A} \textbf{ after } \goLeft{A}, \At{A,1} & \\
\label{sumo:dynamic:4}&\Out{A} \textbf{ after } \goRight{A}, \At{A,l} & \\
\label{sumo:dynamic:5}&\textbf{nonexecutable } \goLeft{A} \textbf{ if } \Out{A} &  \\
\label{sumo:dynamic:6}&\textbf{nonexecutable } \goRight{A} \textbf{ if } \Out{A} &  \\
\label{sumo:dynamic:7}&\textbf{inertial } \At{A,L}, \Out{A} &   
\end{align}

Instantiating this action description with an agent $\boldsymbol{a}$ moving in a two-slot ring where $l=2$
yields a transition system with three states; 
$s_1=\{    \At{\mathbf{a},1},\neg\At{\mathbf{a},2},\neg\Out{\mathbf{a}}\}$,
$s_2=\{\neg\At{\mathbf{a},1},    \At{\mathbf{a},2},\neg\Out{\mathbf{a}}\}$, and
$s_3=\{\neg\At{\mathbf{a},1},\neg\At{\mathbf{a},2},    \Out{\mathbf{a}}\}$,
along with four transitions
$\langle s_1,\{\goRight{\mathbf{a}}\},s_2\rangle$,
$\langle s_2,\{\goLeft{\mathbf{a}}\},s_1\rangle$,
$\langle s_1,\{\goLeft{\mathbf{a}}\},s_3\rangle$, and
$\langle s_2,\{\goRight{\mathbf{a}}\},s_3\rangle$.



\section{Foundations}
\label{sec:foundation}

We consider a set $\boldsymbol{A}$ of agents whose actions are governed by an environment. 
We formalize such multi-agent systems by means of action descriptions in $\mathcal{BC}$,
one for each agent in $\boldsymbol{A}$ and 
descriptions capturing their interplay.

Our formalization has two stages,
which will be described in the following two subsections.
The first one focuses on designating aspects of the domain that do not show up from a single-agent perspective 
but arise once there are multiple agents.
In the second stage, all these aspects are handled so that we get a global view of the overall multi-agent system.

To be more precise, 
the signature of a multi-agent system $(\boldsymbol{A},\boldsymbol{c},\boldsymbol{r})$ consists of actions $\mathcal{A}$ and fluents $\mathcal{F}$.
Each agent $\boldsymbol{a}\in\boldsymbol{A}$ is represented by 
an action description $\mathcal{D}_{\boldsymbol{a}}$ over 
$\mathcal{A}_{\boldsymbol{a}}\subseteq\mathcal{A}$ and 
$\mathcal{F}_{\boldsymbol{a}}\subseteq\mathcal{F}$ such that
$\mathcal{A}_{\boldsymbol{a}}\cap\mathcal{A}_{\boldsymbol{a'}} = \emptyset$ for all $\boldsymbol{a}\neq\boldsymbol{a'}$.
That is, while agents may share fluents, their actions are distinct.
Each action description represents the correct behavior of an agent within the environment from a single-agent perspective 
--- specific agents may or may not behave accordingly.
Components $\boldsymbol{c}$ and $\boldsymbol{r}$ are represented by action descriptions $\mathcal{D}_{\boldsymbol{c}}$ and $\mathcal{D}_{\boldsymbol{r}}$
over $\mathcal{A}$ and $\mathcal{F}$, respectively.  
They correspond to the first and second stage in our methodology respectively.
While the role of the component $\boldsymbol{c}$ is identifying potential conflicts,
the role of $\boldsymbol{r}$ is resolving such conflicts.

\subsection{Identifying potential conflicts (the first stage)} 

Let us extend our previous example with a second Sumo $\boldsymbol{b}$.
This results in two action descriptions $\mathcal{D}_{\boldsymbol{a}}$ and $\mathcal{D}_{\boldsymbol{b}}$,
which are composed of laws~\eqref{sumo:static:1}--\eqref{sumo:dynamic:7} only differing in the used agent identifier,
viz.  $\boldsymbol{a}$ and $\boldsymbol{b}$, respectively.
Both Sumos can thus move left or right to adjacent slots in the ring and drop out at each end.
Also,
there can only be one Sumo in a slot at a time.
A Sumo who moved or is pushed out of the ring cannot return.
It can, however, resist a moving opponent by moving in the reverse direction.
This results in both Sumos staying in their previous slots. 
Similarly, if both Sumos want to move to the same slot at the same time,
they bounce back and stay in their previous slots.

The mere union of all action descriptions capturing single agents in $\boldsymbol{A}$ may lack some valid states of the multi-agent system.
One role of component $\boldsymbol{c}$ is to rectify this via action description $\mathcal{D}_{\boldsymbol{c}}$.
In particular, it may be necessary for the action description $\mathcal{D}_{\boldsymbol{c}}$ to defeat some of the static 
laws of single agent descriptions.
To this end, we turn static laws of single agent descriptions into default rules by introducing abnormality fluents.
\begin{definition}
Let $\mathcal{D}$ be an action description in $\mathcal{BC}$.
Then, we define $\tau(\mathcal{D})$ as the result of
\begin{enumerate}
\item replacing each static  law of form \eqref{eq:static:law} in $\mathcal{D}$ by
\begin{equation}\label{eq:static:ab}
a_0\textbf{ if }a_1,\dots,a_m\textbf{ ifcons } a_{m+1},\dots,a_n,\neg\Ab{a_0}
\end{equation}
\item adding for each defined fluent $\Ab{a_0}$ introduced in \eqref{eq:static:ab} a rule of form
\begin{equation}\label{eq:default:ab}
\textbf{default }\neg\Ab{a_0}
\end{equation}
\item copying each dynamic law without change.
\end{enumerate}
\end{definition}
At the end of the first stage of our formalization of the global view of a multi-agent system,
we get the intermediate action description:
\[
\mathcal{U}_{(\boldsymbol{A},\boldsymbol{c})} = \bigcup_{a \in \boldsymbol{A}} \tau(\mathcal{D}_{\boldsymbol{a}}) \;
      \cup \mathcal{D}_{\boldsymbol{c}}
\]
In our Sumo example, we do not need to defeat any static laws 
since the union of all single agent descriptions 
does not lack any states of the multi-agent system.
In Section~\ref{subsec:defeat:static:law} we give an example domain where the component $\boldsymbol{c}$ has to defeat 
some static laws to generate some previously lacking valid states of $\mathcal{U}_{(\boldsymbol{A},\boldsymbol{c})}$.

\begin{figure}
\caption{A part of the transition system of the union of action descriptions.
Self loops with $\emptyset$ compound action are omitted for clearance.
$l_x$ and $r_x$ abbreviate $\goLeft{x}$ and $\goRight{x}$.}
\begin{center}
\begin{tikzpicture}[->,>=stealth',node distance=3cm]
\tikzstyle{every state}=[scale=0.75]

\node[regular polygon, regular polygon sides=3,minimum width=6cm] (t) {};

\node[state] at (t.center) (s1)  [label={[scale=0.9]10:$s_1$}]     {
\begin{tabular}{|p{0.15cm}|p{0.15cm}|p{0.15cm}|p{0.15cm}|}
\hline
$a$ & & & $b$\\
\hline
\end{tabular}
};

\node[state] at (t.corner 1) (s2) [label={[scale=0.9]60:$s_2$}]   {
\begin{tabular}{|p{0.15cm}|p{0.15cm}|p{0.15cm}|p{0.15cm}|}
\hline
 &$a$&$b$& \\
\hline
\end{tabular}
};

\node[state] at (t.corner 2) (s3)  [label={[scale=0.9]225:$s_3$}]     {
\begin{tabular}{|p{0.15cm}|p{0.15cm}|p{0.15cm}|p{0.15cm}|}
\hline
$a$&&$b$& \\
\hline
\end{tabular}
};

\node[state] at (t.corner 3) (s4)  [label={[scale=0.9]315:$s_4$}]       {
\begin{tabular}{|p{0.15cm}|p{0.15cm}|p{0.15cm}|p{0.15cm}|}
\hline
 &$a$& &$b$\\
\hline
\end{tabular}
};

\node[state] (s5) [left of=s2,node distance=7cm,label={[scale=0.9]135:$s_5$}]  {
\begin{tabular}{|p{0.15cm}|p{0.15cm}|p{0.15cm}|p{0.15cm}|}
\hline
$a$&$b$& & \\
\hline
\end{tabular}
};

\node[state] (s6) [right of=s2,node distance=7cm,label={[scale=0.9]45:$s_6$}]  {
\begin{tabular}{|p{0.15cm}|p{0.15cm}|p{0.15cm}|p{0.15cm}|}
\hline
&& $a$& $b$\\
\hline
\end{tabular}
};

\path (s1.225-10) edge [bend right] node[above,scale=0.9] {$l_b$} (s3.45+10);
\path (s3.45-10) edge [bend right] node[below,scale=0.9] {$r_b$} (s1.225+10);

\path (s1.315-10) edge [bend right] node[below,scale=0.9] {$r_a$} (s4.145+10);
\path (s4.135-10) edge [bend right] node[above,scale=0.9] {$l_a$} (s1.315+10);

\path (s1.90-10) edge [bend right] node[right,scale=0.9] {$r_a,l_b$} (s2.270+10);
\path (s2.270-10) edge [bend right] node[left,scale=0.9] {$l_a,r_b$} (s1.90+10);

\path (s2.225-30) edge [bend right=15] node[left,scale=0.9] {$l_a$} (s3.90+30);
\path (s3.90+20) edge [bend right=15] node[right,scale=0.9] {$r_a$} (s2.225-20);

\path (s2.315+30) edge [bend left=15] node[right,scale=0.9] {$r_b$} (s4.90-30);
\path (s4.90-20) edge [bend left=15] node[left,scale=0.9] {$l_b$} (s2.315+20);

\path (s3.0-30) edge [bend right=15] node[below,scale=0.9] {$r_a,r_b$} (s4.180+30);
\path (s4.180+20) edge [bend right=15] node[above,scale=0.9] {$l_a,l_b$} (s3.0-20);

\path (s3.135+10) edge [bend left=15] node[left,scale=0.9] {$l_b$} (s5.270-10);
\path (s5.270+0) edge [bend left=15] node[right,scale=0.9] {$r_b$} (s3.135-0);

\path (s5.0+20) edge [bend left=15] node[above,scale=0.9] {$r_a,r_b$} (s2.180-20);
\path (s2.180+0) edge [bend left=15] node[below,scale=0.9] {$l_a,l_b$} (s5.0-0);

\path (s6.180-20) edge [bend right=15] node[above,scale=0.9] {$l_a,l_b$} (s2.0+20);
\path (s2.0-0) edge [bend right=15] node[below,scale=0.9] {$r_a,r_b$} (s6.180-0);

\path (s4.45-10) edge [bend right=15] node[right,scale=0.9] {$r_a$} (s6.270+10);
\path (s6.270-0) edge [bend right=15] node[left,scale=0.9] {$l_a$} (s4.45+0);

\end{tikzpicture}
\end{center}
\label{fig:trsys_sumounion}\label{fig:sumo:union:transition:system}
\end{figure}
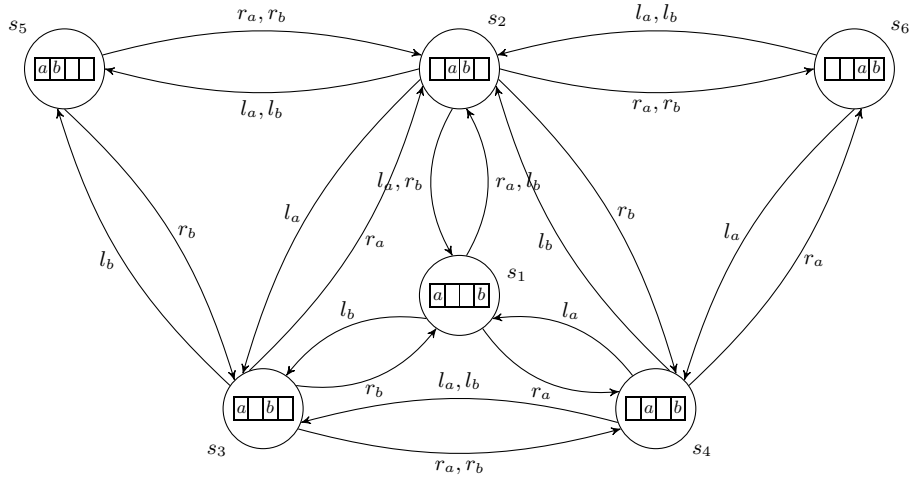

Apart from lacking states,
the mere union of agent action descriptions is prone to invalid states and transitions from a multi-agent perspective.
For instance, the union of action descriptions $\mathcal{D}_{\boldsymbol{a}}$ and $\mathcal{D}_{\boldsymbol{b}}$ tolerates both Sumos in the same slots,
as manifested by the states
\(
\{\At{\boldsymbol{a},L},\At{\boldsymbol{b},L}\} 
\).
%
Also, it permits both Sumos passing through each other, exhibited by the transition
\(
\langle\{\At{a,2},\At{b,3}\},\{\goRight{\boldsymbol{a}},\goLeft{\boldsymbol{b}}\},\{\At{a,3},\At{b,2}\}\rangle .
\)
%
Such invalid states and transitions must be ruled out by appropriate laws in $\mathcal{D}_{\boldsymbol{c}}$ 
to govern the interplay of $\mathcal{D}_{\boldsymbol{a}}$ and $\mathcal{D}_{\boldsymbol{b}}$.
%
Consider the action description $\mathcal{D}_{\boldsymbol{c}}$ consisting of the laws in \eqref{sumo:static:env:1} and \eqref{sumo:dynamic:env:1}.
\begin{align}
\label{sumo:static:env:1} &\neg \At{A,L} \textbf{ if } \At{A',L} & (A \neq A')\\ 
\label{sumo:dynamic:env:1}&\textbf{nonexecutable } \goRight{A}, \goLeft{A'} \textbf{ if } \At{A,L}, \At{A',L+1} & (A \neq A')
\end{align}
The action description $\mathcal{U}_{(\boldsymbol{A},\boldsymbol{c})}$ induces 21 states.
The part of the transition system where Sumo $\boldsymbol{a}$ is left of $\boldsymbol{b}$ is given in Figure~\ref{fig:sumo:union:transition:system}.

Note that $ab$ fluents introduced by the transformation $\tau$ are statically defined
and they are false by default due to law \eqref{eq:default:ab}.
Since there are no laws in $\mathcal{D}_{\boldsymbol{c}}$ causing an $ab$ fluent to be true in the Sumo example,
there are no states of $\mathcal{U}_{(\boldsymbol{A},\boldsymbol{c})}$ in which an $ab$ fluent holds.

The other role of component $\boldsymbol{c}$,
as stated at the beginning of this section,
is to designate aspects of the domain regarding the interplay of multiple agents in the environment.
To this end, we first define the notion of \emph{potential conflict} to cover such multi-agent aspects.
A potential conflict is a compound action that cannot be executed in a state in view of the laws of an action description.
%
\begin{definition}\label{def:conflict}
Let $\mathcal{D}$ be an action description in $\mathcal{BC}$,
$(S(\mathcal{D}),T(\mathcal{D}))$ the corresponding transition system,
and $s\in S(\mathcal{D})$.

We define a potential conflict in $s$ and $\mathcal{D}$ as
\[
\raisebox{2pt}{$\chi$}_{\mathcal{D}}(s)
=
\{
c \subseteq \mathcal{A} \mid \textit{there exists no } s' \in S(\mathcal{D}) \textit{ s.t. } \langle s,c,s'\rangle \in T(\mathcal{D})
\}
\,.
\]
\end{definition}

In our running example, there are basically 3 cases of such multi-agent aspects of the domain:
(i) a Sumo can push another Sumo,
(ii) Sumos can resist a push by moving in the reverse direction, and
(iii) two Sumos bounce back when they want to move to the same slot at the same time.
The description $\mathcal{D}_{\boldsymbol{c}}$ composed of laws \eqref{sumo:static:env:1} and \eqref{sumo:dynamic:env:1}
already identifies all potential conflicts corresponding to these 3 cases.
%
For some representative potential conflicts of (ii) and (iii) case,
consider the compound action $\{\goRight{\boldsymbol{a}},\goLeft{\boldsymbol{b}}\}$.
While it can be used to switch from $s_1$ to $s_2$,
it cannot be used to leave any other state in Figure~\ref{fig:sumo:union:transition:system}.
A transition with this compound action is ruled out in states $s_5$, $s_2$, and $s_6$ by \eqref{sumo:dynamic:env:1}
and in $s_3$ and $s_4$ by \eqref{sumo:static:env:1}.
Clearly, we have
\(
\{\goRight{\boldsymbol{a}},\goLeft{\boldsymbol{b}}\}\in\raisebox{2pt}{$\chi$}_{\mathcal{U}_{(\{\boldsymbol{a},\boldsymbol{b}\},\boldsymbol{c})}}(s)
\)
for $s=s_2,\dots,s_6$.
Additionally, for the first case,
we observe that
\(
\{\goRight{\boldsymbol{a}}\}\in\raisebox{2pt}{$\chi$}_{\mathcal{U}_{(\{\boldsymbol{a},\boldsymbol{b}\},\boldsymbol{c})}}(s)
\)
and
\(
\{\goLeft{\boldsymbol{b}}\}\in\raisebox{2pt}{$\chi$}_{\mathcal{U}_{(\{\boldsymbol{a},\boldsymbol{b}\},\boldsymbol{c})}}(s)
\)
for $s=s_2,s_5,s_6$.


\subsection{Resolving potential conflicts (the second stage)}  

The first stage of our formalization ends with identifying potential conflicts. 
Such conflicts are inadmissible in a multi-agent setting.
It should be allowed for the individual agents to perform the corresponding compound action.
The second stage is about preventing identified conflicts that are related to the multi-agent aspects of the domain
from becoming actual conflicts.
The role of the conflict resolution component $\boldsymbol{r}$ 
in the multi-agent system $(\boldsymbol{A},\boldsymbol{c},\boldsymbol{r})$ is to rectify this.
To this end, the action description $\mathcal{D}_{\boldsymbol{r}}$ has to defeat some of the dynamic laws of
$\mathcal{U}_{(\boldsymbol{A},\boldsymbol{c})}$.

Similar to the transformation $\tau$,
we turn dynamic laws of the description $\mathcal{U}_{(\boldsymbol{A},\boldsymbol{c})}$ 
into default rules by introducing an abnormality fluent using the transformation $\beta$.
\begin{definition}
Let $\mathcal{D}$ be an action description in $\mathcal{BC}$.

Then, we define $\beta(\mathcal{D})$ as the result of
\begin{enumerate}
\item replacing each dynamic  law of form \eqref{eq:dynamic:law} in $\mathcal{D}$ by
\begin{equation}\label{eq:dynamic:ab}
a_0\textbf{ after }a_1,\dots,a_m\textbf{ ifcons } a_{m+1},\dots,a_n,\neg\Abd{a_0}
\end{equation}
\item adding for each fluent $\Abd{a_0}$ introduced in \eqref{eq:dynamic:ab} a rule of form
\begin{equation}\label{eq:default:ab:beta}
\textbf{default }\neg\Abd{a_0}
\end{equation}
\item copying each static law without change.
\end{enumerate}
\end{definition}
Note that unlike transformation $\tau$,
abnormality fluents introduced in $\beta$ are regular fluents.
This allows the modeler to use such fluents in heads of dynamic laws in $\mathcal{D}_{\boldsymbol{r}}$.
At the end of the second stage of the formalization we get 
the global view of the  overall multi-agent system represented by the action description:
\[
\mathcal{M}_{(\boldsymbol{A},\boldsymbol{c},\boldsymbol{r})} = \beta(\mathcal{U}_{(\boldsymbol{A},\boldsymbol{c})}) 
      \cup \mathcal{D}_{\boldsymbol{r}}
\]


Before illustrating the resolution step for our running example,
we analyze the properties of transformation $\beta$ and action description $\mathcal{D}_{\boldsymbol{r}}$.
%
%
Due to Lemma~\ref{lemma:beta:conflicts}, 
we can be sure that potential conflicts of $\mathcal{U}_{(\boldsymbol{A},\boldsymbol{c})}$ are preserved after applying $\beta$.

\begin{lemma} 
\label{lemma:beta:conflicts}
Let $\mathcal{D}$ be an action description in $\mathcal{BC}$.
$c \in \raisebox{2pt}{$\chi$}_{\mathcal{D}}(s)$ iff
$c \in \raisebox{2pt}{$\chi$}_{\mathcal{\beta(D)}}(s^{*})$ 
where $s \subseteq s^*$ and $s^* \setminus s$ has either $ab'$ or $\neg ab'$ for each $ab'$ fluent introduced in $\beta(\mathcal{D})$ and no other literal.
\end{lemma}

Although the modeler knows that potential conflicts identified in the first stage of the formalization are preserved by $\beta$,
she is not guided on the structure of laws needed in $\mathcal{D}_{\boldsymbol{r}}$ in order to resolve a conflict.
To remedy this situation, we define a condition of a dynamic law being \emph{covered by a compound action at a state}.

\begin{definition}
A dynamic law of the form \eqref{eq:dynamic:law} in $\mathcal{D}$ is covered by compound action $c$ at state $s \in S(\mathcal{D})$ 
iff for all $1 \leq i \leq m$
(i) $a_i \in c$ when $a_i$ is a positive action literal,
(ii) $\overline{a_i} \not\in c$ when $a_i$ is a negative action literal, and
(iii) $a_i \in s$ when $a_i$ is a fluent literal.
\end{definition}

Furthermore, Lemma~\ref{lemma:beta:coveredlaws} shows that laws of $\mathcal{D}_{\boldsymbol{r}}$ needed for resolving a potential conflict
must be dynamic laws that are covered by the related compound action and the state.
When the compound action and state are clear from the context,
we may only use the statement that a law is covered.

\begin{lemma}
\label{lemma:beta:coveredlaws}
Let $\mathcal{D}$ be an action description in $\mathcal{BC}$ over actions $\mathcal{A}$.
Given a compound action $c \subseteq \mathcal{A}$ and a state $s \in S(\mathcal{D})$,
$\langle s,c,s' \rangle \in T(\mathcal{D})$ iff 
$\langle s,c,s' \rangle \in T(\mathcal{D'})$ 
where $\mathcal{D'}$ is equal to $\mathcal{D}$ except all its dynamic laws not covered by $c$ at $s$ are taken out.
\end{lemma}

For resolving a potential conflict, 
the modeler has to encode covered dynamic laws in $\mathcal{D}_{\boldsymbol{r}}$.
Basically, some laws in $\mathcal{D}_{\boldsymbol{r}}$ should first defeat dynamic laws of $\beta(\mathcal{U}_{(\boldsymbol{A},\boldsymbol{c})})$ causing
contradictory effects w.r.t. the desired successor state .
Then additional laws in $\mathcal{D}_{\boldsymbol{r}}$ generate effect fluent literals that are previously not possible.
We illustrate this methodology using our running example.

Consider the potential conflict related to two Sumos trying to move to the same slot at the same time for the state $s_4$ in Figure~\ref{fig:sumo:union:transition:system},
i.e., $\{\goRight{\boldsymbol{a}},\goLeft{\boldsymbol{b}}\}\in\raisebox{2pt}{$\chi$}_{\mathcal{U}_{(\{\boldsymbol{a},\boldsymbol{b}\},\boldsymbol{c})}}(s_4)$.
Considering instances of dynamic laws \eqref{sumo:dynamic:1}, \eqref{sumo:dynamic:2}, and \eqref{sumo:dynamic:7}, 
the laws \eqref{law:resolve:1}--\eqref{law:resolve:4} in $\beta(\mathcal{U}_{(\boldsymbol{A},\boldsymbol{c})})$ are covered by $\{\goRight{\boldsymbol{a}},\goLeft{\boldsymbol{b}}\}$ at $s_4$.
\begin{align}
\label{law:resolve:1}&\At{a,3} \textbf{ after } \goRight{a}, \At{a,2} \textbf{ ifcons } \neg \Abd{\At{a,3}}  \\
\label{law:resolve:2}&\At{b,3} \textbf{ after } \goLeft{b}, \At{b,4} \textbf{ ifcons } \neg \Abd{\At{b,3}}  \\
\label{law:resolve:3}&\At{a,2} \textbf{ after } \At{a,2} \textbf{ ifcons } \At{a,2}, \neg \Abd{\At{a,2}} \\
\label{law:resolve:4}&\At{b,4} \textbf{ after } \At{a,4} \textbf{ ifcons } \At{b,4}, \neg \Abd{\At{b,4}} 
\end{align}
Since Sumos bounce back in this case, 
effect fluents $\At{a,3}$ and $\At{b,3}$ do not hold in the desired successor state.
Hence, we need to defeat the laws \eqref{law:resolve:1} and \eqref{law:resolve:2} by causing abnormality fluents $\Abd{\At{a,3}}$ and $\Abd{\At{b,3}}$
to be true in order to resolve the potential conflict.
In the general case, laws \eqref{law:resolve:bounce:1} and \eqref{law:resolve:bounce:2} in $\mathcal{D}_{\boldsymbol{r}}$ 
cause these abnormality fluents and 
resolve all potential conflicts related to Sumos trying to move to the same slot at the same time.
Considering these potential conflicts,
notice that laws \eqref{law:resolve:bounce:1} and \eqref{law:resolve:bounce:2} are covered by respective compound actions at respective states
as formally stated in Lemma~\ref{lemma:beta:coveredlaws}.
\begin{align}
&\Abd{\At{A,L+1}} \textbf{ after } \goRight{A}, \goLeft{A'}, \At{A,L}, \At{A',L+2}, \nonumber \\
\label{law:resolve:bounce:1}&\qquad\qquad\qquad\qquad\neg \goLeft{A}, \neg \goRight{A'} \qquad\qquad(A \neq A') \\
&\Abd{\At{A',L+1}} \textbf{ after } \goRight{A}, \goLeft{A'}, \At{A,L}, \At{A',L+2}, \nonumber \\
\label{law:resolve:bounce:2}&\qquad\qquad\qquad\qquad\neg \goLeft{A}, \neg \goRight{A'} \qquad\qquad(A \neq A') 
\end{align}

Next, we resolve the potential conflicts related to a Sumo pushing another Sumo.
For example, consider 
$\{\goRight{\boldsymbol{a}}\}\in\raisebox{2pt}{$\chi$}_{\mathcal{U}_{(\{\boldsymbol{a},\boldsymbol{b}\},\boldsymbol{c})}}(s_2)$.
In such a case, at the desired successor state
the pushed Sumo moves one slot in the direction of push
(and maybe pushed out of the ring if he is at a border slot).
Dynamic laws \eqref{law:resolve:push:1}--\eqref{law:resolve:push:4} in $\mathcal{D}_{\boldsymbol{r}}$ resolve such potential conflicts.
\begin{align}
\label{law:resolve:push:1}&\At{A',L+2} \textbf{ after } \At{A,L}, \At{A',L+1}, \goRight{A}, \neg \goLeft{A'}  &&  \nonumber\\
&\qquad\qquad\qquad  (A \neq A', L+1 < 4) && \\
\label{law:resolve:push:2}&\At{A,L-1} \textbf{ after } \At{A,L}, \At{A',L+1}, \goLeft{A'}, \neg \goRight{A} && \nonumber\\
&\qquad\qquad\qquad (A \neq A', L > 1) &&\\
\label{law:resolve:push:3}&\Out{A'} \textbf{ after } \At{A,L}, \At{A',L+1}, \goRight{A}, \neg \goLeft{A'} && \nonumber\\
&\qquad\qquad\qquad (A \neq A', L = 3) &&\\
\label{law:resolve:push:4}&\Out{A} \textbf{ after } \At{A,L}, \At{A',L+1}, \goLeft{A'}, \neg \goRight{A} && \nonumber\\
&\qquad\qquad\qquad (A \neq A', L = 1) &&
\end{align}
Note that unlike the previous case, we have not defeated some covered laws by explicitly causing some $ab'$ fluents to be true.
Actually, related literals in the successor state caused by laws \eqref{law:resolve:push:1}--\eqref{law:resolve:push:4} defeat 
contradictory covered laws (in this case laws of intertia for the pushed Sumo) via indirect effects.

The last case concerns the potential conflicts related to two Sumos trying to push each other at the same time.
In the successor state Sumos should be in their previous slots.
The laws \eqref{law:resolve:resist:1}--\eqref{law:resolve:resist:3} in $\mathcal{D}_{\boldsymbol{r}}$ defeat contradictory covered laws,
which represent effects of movement, 
by causing related $ab'$ fluents to be true.
\begin{align}
&\Abd{\At{A,L+1}} \textbf{ after } \goRight{A}, \goLeft{A'}, \At{A,L}, \At{A',L+1}, \nonumber \\
\label{law:resolve:resist:1}&\qquad\qquad\qquad\qquad\neg \goLeft{A}, \neg \goRight{A'} \qquad\qquad(A \neq A') \\
&\Abd{\At{A',L}} \textbf{ after } \goRight{A}, \goLeft{A'}, \At{A,L}, \At{A',L+1}, \nonumber \\
\label{law:resolve:resist:2}&\qquad\qquad\qquad\qquad\neg \goLeft{A}, \neg \goRight{A'} \qquad\qquad(A \neq A') \\
&\Abd{n(A,A',L)} \textbf{ after } \goRight{A}, \goLeft{A'}, \At{A,L}, \At{A',L+1}, \nonumber \\
\label{law:resolve:resist:3}&\qquad\qquad\qquad\qquad\neg \goLeft{A}, \neg \goRight{A'} \qquad\qquad(A \neq A') 
\end{align}

The law \eqref{law:resolve:resist:3} defeats the law in $\beta(\mathcal{U}_{(\boldsymbol{A},\boldsymbol{c})})$ 
that is $\beta$ transformed version of \eqref{sumo:dynamic:env:1}.
In principle such nonexecutable laws are abbreviations of pairs of dynamic laws with contradictory head fluents \cite{leliya13a}.
We assume that these head literals are unique for each nonexecutable law
in order to avoid overly defeating such laws in the second stage.
For an action description, this can be achieved by introducing some fresh fluents for each of such dynamic law pairs.
Otherwise, a slightly modified version of $\beta$ may generate a unique id in the introduced abnormality fluent
for each nonexecutable law.\footnote{This also applies to impossible laws and the transformation $\tau$.}
For instance, in our Sumo example $\beta$ adds the $\neg \Abd{n(A,A',L)}$ abnormality fluent to the ifcons part 
when transforming dynamic laws abbreviated by \eqref{sumo:dynamic:env:1} in a uniquely identified fashion.

Proposition~\ref{prop:resolutionguarantee} guarantees that a potential conflict in $\mathcal{U}_{(\boldsymbol{A},\boldsymbol{c})}$ can 
always be resolved in $\mathcal{M}_{(\boldsymbol{A},\boldsymbol{c},\boldsymbol{r})}$.
The resolution used in the proposition is clearly not the only way 
and may also not be the pragmatic way to resolve a potential conflict
(for instance, a more concise way has already been illustrated in the Sumo example).
However, it forms a basic guideline as a general methodology on resolving conflicts,
i.e., defeating dynamic laws causing contradictory effects and generating desired effects in the successor state.

\begin{proposition}
\label{prop:resolutionguarantee}
Let $\mathcal{D}$ be an action description in $\mathcal{BC}$ over actions $\mathcal{A}$ and 
$c \subseteq \mathcal{A}$ be a compound action.
If $c \in \raisebox{2pt}{$\chi$}_{\mathcal{D}}(s)$,
then for any state $s' \in S(\mathcal{D})$, 
$\langle s^*,c,s' \cup d \rangle \in T(\beta(\mathcal{D}) \cup \mathcal{R})$ such that
$s \subseteq s^*$ and $s^* \setminus s$ has either $ab'$ or $\neg ab'$ for each $ab'$ fluent introduced in $\beta(\mathcal{D})$ and no other literal;
$\mathcal{R}$ is a set of dynamic laws covered by $c$ at $s$ of the form
\begin{equation}\label{eq:resolve:defeat}
\Abd{a_0}\textbf{ after }a_1,\dots,a_m
\end{equation}
for each dynamic law in $\mathcal{D}$ that is covered by $c$ at $s$,
where $\Abd{a_0}$ is introduced in $\beta(\mathcal{D})$ for the covered law in $\mathcal{D}$,
and of the form
\begin{equation}\label{eq:resolve:succstate}
f\textbf{ after }a_1,\dots,a_m
\end{equation}
for each literal $f \in s'$; and
set $d$ is composed of positive $ab'$ literals that appear in heads of laws \eqref{eq:resolve:defeat} in $\mathcal{R}$ and
negative $ab'$ literals for all the rest $ab'$ fluent symbols introduced in $\beta(\mathcal{D})$.
\end{proposition}

Let $\mathcal{D}_{\boldsymbol{r}}$ be composed of laws \eqref{law:resolve:bounce:1}--\eqref{law:resolve:resist:3}.
At the end of the second stage, the action description $\mathcal{M}_{(\boldsymbol{A},\boldsymbol{c},\boldsymbol{r})}$
gives the global view of the Sumo agents.
Considering a specific potential conflict, for instance,
$\{\goRight{\boldsymbol{a}},\goLeft{\boldsymbol{b}}\}\in\raisebox{2pt}{$\chi$}_{\mathcal{U}_{(\{\boldsymbol{a},\boldsymbol{b}\},\boldsymbol{c})}}(s_4)$
holds given the state $s_4$ in Figure~\ref{fig:sumo:union:transition:system}.
After the resolution stage, however,
it is not anymore a potential conflict of $\mathcal{M}_{(\boldsymbol{A},\boldsymbol{c},\boldsymbol{r})}$ and
the transition diagram has a corresponding transition,
i.e., $\langle s_4, \{\goRight{\boldsymbol{a}},\goLeft{\boldsymbol{b}}\}, s_7 \rangle \in T(\mathcal{M}_{(\boldsymbol{A},\boldsymbol{c},\boldsymbol{r})})$ 
where the successor state $s_7 = \{ \At{a,2}, \At{b,4}, \Abd{\At{a,3}}, \Abd{\At{b,3}} \}$
(Sumos bounce back and stay in their previous slots).

Since the $ab'$ fluents introduced by the transformation $\beta$ are regular fluents,
there may be states of $\mathcal{M}_{(\boldsymbol{A},\boldsymbol{c},\boldsymbol{r})}$ that include superfluous $ab'$ fluents.
In our Sumo example, for instance,
$\{ \At{a,1}, \At{b,4}, \Abd{\At{a,3}} \} \in S(\mathcal{M}_{(\boldsymbol{A},\boldsymbol{c},\boldsymbol{r})})$ holds.
Such states, however,
are not accessible from sound initial states.
We plan to address this issue formally by augmenting our formalization with a query language that enables one to express initial states and reasoning tasks
(see also Section~\ref{sec:discussion} for the related future work).

\subsection{Defeating static laws in the first stage} 
\label{subsec:defeat:static:law}

Recall that in the Sumo example, we have not used $ab$ fluents introduced by $\tau$ to defeat some static laws,
since all valid states of the multi-agent domain are generated by the mere union of agent descriptions of single agent perspective.
Consider another domain where there are two agents next to opposite ends of a table \cite{pednault87a}.
The table is initially on the floor.
Each agent may lift the table up using its respective end.
Whenever an agent lifts up the table, 
it holds the table steady in its resulting state.
The table is fully lifted when it is lifted from both ends by respective agents.

Agent $\boldsymbol{l}$ is next to the left end of the table.
The behaviour of this agent can be captured through fluents $\Table{P}$ and
the action $\liftLeft$ where $P \in \lbrace onfloor, leftup \rbrace$.
The fluent $\Table{leftup}$ represents the situation where the left end of the table is lifted up and 
the right end of it is on the floor.

The behaviour of $\boldsymbol{l}$ from a single agent perspective can be modeled by the action description $\mathcal{D}_{\boldsymbol{l}}$
that is composed of laws \eqref{table:static:1}--\eqref{table:dynamic:3}.
\begin{align}
\label{table:static:1}&\neg \Table{leftup} \textbf{ if } \Table{onfloor} & \\ 
\label{table:static:2}&\neg \Table{onfloor} \textbf{ if } \Table{leftup} & \\ 
\label{table:static:3}&\textbf{impossible } \neg \Table{onfloor}, \neg \Table{leftup} & \\
\label{table:dynamic:1}&\Table{leftup} \textbf{ after } \liftLeft  & \\
\label{table:dynamic:2}&\textbf{nonexecutable } \liftLeft \textbf{ if } \Table{leftup} & \\
\label{table:dynamic:3}&\textbf{inertial } \Table{P} & 
\end{align}
$\mathcal{D}_{\boldsymbol{l}}$ has 2 states;
$s_1 = \{ \Table{onfloor}, \neg \Table{leftup} \}$ and 
$s_2 = \{ \Table{leftup},\linebreak \neg \Table{onfloor} \}$.
It has 3 transitions;
$\langle s_1, \{ \}, s_1 \rangle$,
$\langle s_2, \{ \}, s_2 \rangle$,
and $\langle s_1, \{ \liftLeft \}, s_2 \rangle$.

The behaviour of agent $\boldsymbol{r}$ is similar to that of $\boldsymbol{l}$ 
and the description $\mathcal{D}_{\boldsymbol{r}}$ is equal to $\mathcal{D}_{\boldsymbol{l}}$ 
except fluent $\Table{rightup}$ and action $\liftRight$ are used instead of 
$\Table{leftup}$ and $\liftLeft$ respectively.

The mere union of descriptions $\mathcal{D}_{\boldsymbol{l}}$ and $\mathcal{D}_{\boldsymbol{r}}$ has two states;
a state where the only positive fluent literal is $\Table{onfloor}$ 
and an invalid state where both $\Table{leftup}$ and $\Table{rightup}$ hold.
Besides, it is easy to see that the state where the table is fully lifted 
is invisible to agents from a single agent perspective.

At the end of the first stage of our formalization,
$\mathcal{U}_{(\boldsymbol{A},\boldsymbol{c})}$
must cover all valid states of the multi-agent system.
Let $\mathcal{D}_{\boldsymbol{c}}$ be laws \eqref{table:conflict:1}--\eqref{table:conflict:8}.
Note that $\mathcal{D}_{\boldsymbol{c}}$ uses the new fluent $\Table{lifted}$ to cover a state
that is invisible from single agent perspective.
However, this needs defeating static laws \eqref{table:static:3} from descriptions $\mathcal{D}_{\boldsymbol{l}}$ and $\mathcal{D}_{\boldsymbol{r}}$.
This is achieved by static laws \eqref{table:conflict:4} and \eqref{table:conflict:5} where 
$\neg \Ab{imp(l)}$ and $\neg \Ab{imp(r)}$ are abnormality fluents introduced by $\tau$ to make the corresponding impossible laws defeasible.
\begin{align}
\label{table:conflict:1}&\neg \Table{P} \textbf{ if } \Table{lifted} \qquad\qquad\qquad\qquad (P \neq lifted) \\ 
\label{table:conflict:2}&\neg \Table{lifted} \textbf{ if } \Table{P} \qquad\qquad\qquad\qquad (P \neq lifted) \\ 
\label{table:conflict:3}&\neg \Table{P} \textbf{ if } \Table{P'}  \quad (P \neq P'; P,P' \in \{rightup,leftup\}) \\ 
\label{table:conflict:4}&\Ab{imp(l)}  \\
\label{table:conflict:5}&\Ab{imp(r)}  \\
\label{table:conflict:6}&\textbf{impossible } \neg \Table{onfloor}, \neg \Table{leftup}, \neg \Table{rightup}, \neg \Table{lifted}  \\
\label{table:conflict:7}&\textbf{nonexecutable } \liftLeft \textbf{ if } \Table{P}  \quad (P \not\in \{leftup,onfloor\}) \\
\label{table:conflict:8}&\textbf{nonexecutable } \liftRight \textbf{ if } \Table{P} \quad (P \not\in \{rightup,onfloor\}) 
\end{align}

Dynamic laws \eqref{table:conflict:7} and \eqref{table:conflict:8} eliminate some invalid transitions so that
all multi-agent aspects of the domain are identified by potential conflicts in $\mathcal{U}_{(\boldsymbol{A},\boldsymbol{c})}$.
There are two cases of multi-agent aspects of the domain; 
the table is fully lifted when both agents lift from their respective ends at the same time, or
when an agent lifts up the table from his end and the table is already lifted up from the opposite end.

$\mathcal{U}_{(\boldsymbol{A},\boldsymbol{c})}$ has 4 states,
one for each position of the table.
All the states satisfy literals $\Ab{imp(l)}$ and $\Ab{imp(r)}$.
It has 6 transitions.
Among the 6 transitions, 4 are from each state to itself with no performed actions.
The remaining transitions are from $\{ \Table{onfloor} \}$ 
to $\{ \Table{leftup} \}$ and
$\{ \Table{rightup} \}$ by
compound actions  $\{ \liftLeft \}$ and $\{ \liftRight \}$, respectively.\footnote{\label{fnote:posstate}%
Here we show only positive fluent literals satisfied in a state and omit the literals $\Ab{imp(l)}$ and $\Ab{imp(r)}$.}

One important remark is that the modeler may encode $\mathcal{U}_{(\boldsymbol{A},\boldsymbol{c})}$ in a more compact way.
Our formalization is independent of how it is encoded.
Given that $\mathcal{U}_{(\boldsymbol{A},\boldsymbol{c})}$ covers all valid states of the multi-agent domain
and identifies potential conflicts for all desired multi-agent aspects of the domain,
it can be used in the second stage,
where potential conflicts are resolved and
the global view is captured by $\mathcal{M}_{(\boldsymbol{A},\boldsymbol{c},\boldsymbol{r})}$.

For the second stage of our formalization,
we only show resolution of potential conflict related to the case which is about
both agents lifting from their respective ends at the same time.
Observe that
$\{\liftLeft, \liftRight \}\in\raisebox{2pt}{$\chi$}_{\mathcal{U}_{(\boldsymbol{A},\boldsymbol{c})}}(s)$ 
where $s = \{ \Table{onfloor} \}$.\footnotemark[\getrefnumber{fnote:posstate}]
The dynamic laws \eqref{table:resolve:1}--\eqref{table:resolve:3} in $\mathcal{D}_{\boldsymbol{r}}$ effectively resolve this potential conflict.
While laws \eqref{table:resolve:2} and \eqref{table:resolve:3} defeat the laws in $\beta(\mathcal{U}_{(\boldsymbol{A},\boldsymbol{c})})$
related to individual effects of actions $\liftLeft$ and $\liftRight$,
\eqref{table:resolve:1} caused the effect fluent in the desired successor state.
\begin{align}
\label{table:resolve:1}&\Table{lifted} \textbf{ after } \liftLeft, \liftRight, \Table{onfloor} \\
\label{table:resolve:2}&\Abd{\Table{leftup}} \textbf{ after } \liftLeft, \liftRight, \Table{onfloor} \\
\label{table:resolve:3}&\Abd{\Table{rightup}} \textbf{ after } \liftLeft, \liftRight, \Table{onfloor}
\end{align}



\section{Discussion and Future Work}\label{sec:discussion}

We have developed a formalization for capturing global view of MAS domains.
Methodologically,
we start representing the behaviour of each agent by an action description in $\mathcal{BC}$ from a single agent perspective.
Then,
a two-stage process guides the modeler in composing these single agent descriptions into a single description representing 
the global view of the overall MAS domain.
While the modeler designates multi-agent aspects of the domain via potential conflicts in the first stage,
she resolves these conflicts according to the expected behaviour of the overall system in the second stage.
Considering that representing single agent descriptions is relatively simpler than representing multi-agent description directly,
the formalization developed here is valuable from a knowledge representation perspective
and is different from earlier works using action languages to represent MAS domains in a monolithic way.

Our choice of $\mathcal{BC}$ as the used action language is backed by its clean semantics based on ASP
and ability to express defeasible laws.
Our formalization, however,
is not developed with a fixed action language in mind.
One can use another action language instead and update the methodology with less effort 
given that it can express defeasible laws (e.g., $\mathcal{C}$~\cite{giulif98a}).

The global view of a MAS domain is useful for carrying out reasoning tasks such as projection, planning, or postdiction.
We plan to extend our formalization with a query language so that a modeler can represent such reasoning tasks.
Using the query language, for example,
an initial and goal state of the domain can be represented to facilitate planning for the overall system.

The formalization introduced here paves the way for a number of avenues for future work.
An important one is to utilize online solving capacity of the ASP solver $\mathit{clingo}$~\cite{gekakasc14b}
to develop a complete multi-agent architecture based on logical representations.
The solver may control execution in the environment using the formalization of the global view.
Moreover, 
this can be performed without restarting the solver from scratch every time it communicates with the real agents residing in the environment
with the help of its online solving capacity.


\bibliographystyle{splncs}

\end{document}